\documentclass[12pt]{article} 

\usepackage{ol2}
\usepackage{amsmath,amssymb}

\begin{document}
\title{Negative group velocity and three-wave mixing in  dielectric crystals }
\author{Vitaly V. Slabko,$^{1}$  Sergey A. Myslivets,$^2$  Mikhail I. Shalaev,$^1$ and Alexander K. Popov$^{3,*}$    }
\address{
$^1$Siberian Federal University, 660041 Krasnoyarsk, Russia
\\
$^2$Institute of Physics of the Russian Academy of Sciences, 660036 Krasnoyarsk, Russia
\\
$^3$University of Wisconsin-Stevens Point, Stevens Point, WI 54481, USA\\
$^*$Corresponding author: apopov@uwsp.edu
}
\date{}
\begin{abstract}
Extraordinary features of optical parametric amplification of Stokes electromagnetic waves are investigated, which originate from the three-wave mixing of two ordinary electromagnetic and one backward phonon wave with negative group velocity. A similarity with the counterpart in the negative-index plasmonic metamaterials and differences with those   utilizing contra-propagating ordinary electromagnetic waves as well as electromagnetic and acoustic phonon waves are shown. They stem from  the backwardness of the optical phonons in crystals with negative dispersion.  Nonlinear-optical photonic devices with the properties similar to those predicted for negative-index metamaterials are proposed.
\end{abstract}
%

\sloppy

\section{Introduction}
Optical negative-index (NI) metamaterials (NIMs) form a novel class of electromagnetic media that promises revolutionary breakthroughs in photonics \cite{Sh}. The possibilities of such breakthroughs originate from  backwardness, the extraordinary property that electromagnetic waves acquire in NIMs.  Unlike ordinary positive-index  materials, the energy flow, $\mathbf{S}$,
and the wave-vector, $\mathbf{k}$,
become counter-directed in NIMs, which determines their extraordinary linear and nonlinear optical (NLO) propagation properties. Usually, NIMs are nanostructured metal-insulator composites with a special design of their building blocks at the nanoscale \cite{SM,Zh}. Generally, the metal component  imposes strong absorption of optical radiation in NIMs, which presents a major obstacle towards their numerous exciting prospective applications. Extraordinary features of coherent NLO energy conversion processes in NIMs that stem from wave-mixing of ordinary and backward electromagnetic waves and the possibilities to apply them for compensating the outlined losses  have been shown in \cite{OL,EPJD}. Essentially different properties of three-wave mixing (TWM) and second harmonic generation have been revealed in \cite{LPL,APB1}. Herein, we propose and investigate a different scheme of TWM of ordinary and backward waves (BW). It builds on the stimulated Raman scattering (SRS) where two ordinary electromagnetic waves excite backward elastic vibrational wave in a crystal, which results in TWM. The possibility of such BWs was predicted by L. I. Mandelstam in 1945 \cite{Ma}, who also had pointed out that negative refraction is a general property of the BWs. This letter is to show the possibility to replace negative index composites with the extensively studied ordinary crystals and thus to simulate  unparallel properties of coherent NLO energy exchange between the ordinary and backward waves.  We show  extraordinary nonlinear propagation and output properties of the Stokes electromagnetic wave in one of two different coupling geometries, both utilizing backward elastic waves. The possibility to localize coherent energy conversion and to fit it in to the crystal of given thickness is shown. Such unusual properties are in a striking contrast with those attributed to the counterparts in the standard schemes that build on the coupling of co-propagating photons and phonons \cite{ShB,Boy}. They are also different from  the properties of the phase-matched mixing of optical and acoustic waves for the case when the latter has  energy flux and wave vector directed against those of one of the optical waves \cite{Bob}. The revealed properties can be utilized for  creation of optical switches, filters, amplifiers  and cavity-free optical parametric oscillators based on ordinary Raman crystals without the requirement of periodically poling at the nanoscale \cite{Kh} (and references therein).
\section{Optical phonons, negative group velocity and phase matching options}
\begin{figure}[h]
\begin{center}
\includegraphics[width=.98\columnwidth]{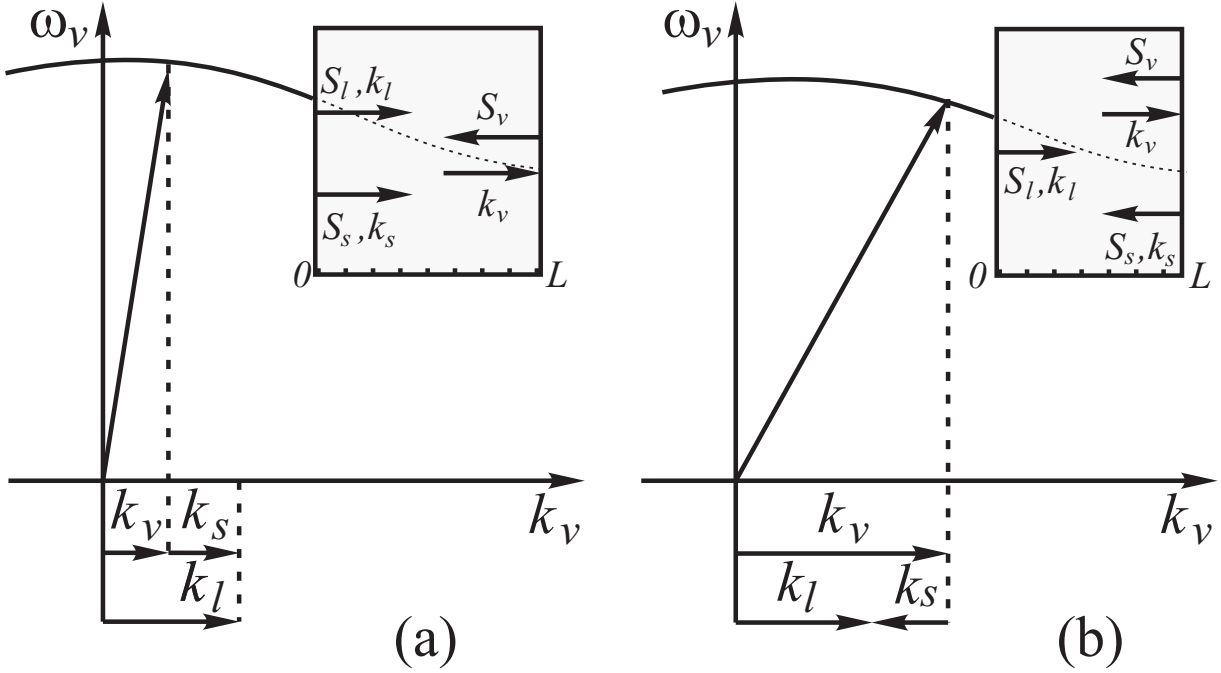}
\caption{\label{f1} Negative dispersion of optical phonons and two phase matching options: (a) -- co-propagating, (b) -- contra-propagating  fundamental, control,  and Stokes, signal, waves. Insets: relative directions of the energy flows and the wave-vectors.  }
\end{center}
\end{figure}
The dispersion curve $\omega(k)$ of phonons in the crystals containing more than one atom per unit cell has two branches: acoustic and optical. For the optical branch, the dispersion is negative,  in the range from zero to the boundary of the first Brillouin's zone (Fig. \ref{f1}) and the group velocity of optical phonons, $\mathbf{v}_{v}^{gr}$, is antiparallel with respect to its wave-vector, $\mathbf{k}_{v}^{ph}$, and phase velocity, $\mathbf{v}_{v}^{ph}$, because $v_{gr}=\partial \omega(k)/\partial k < 0$. Optical vibrations can be excited by the light waves due to the two-photon (Raman) scattering. The latter gives the ground to consider crystal as the analog of the medium with negative refractive index at the phonon frequency and to examine the processes of parametric interaction of the three waves, two of which are ordinary electromagnetic  waves and the third is the wave of elastic vibrations with the directions of the energy flow and of the wave-vector opposite to each other.
\section{Basic equations}
We will  examine only lowest-order Raman process \cite{ShB,Boy}. The interacting waves are given by the equations
\begin{eqnarray*} \label{eq1}
 E_{l,s} &=&({1}/{2})\varepsilon _{l,s} (z,t)e^{ik_{l,s} z-i\omega _{l,s}
t}+c.c. ,\\
 Q_v &=&({1}/{2})Q(z,t)e^{ik_v z-i\omega _v t}+c.c.
\end{eqnarray*}
Here, $\varepsilon _{l,s} $,  $Q$, $\omega _{l,s,v} $ and $k_{l,s,v} $ are the amplitudes, frequencies and wave-vectors of the  fundamental, Stokes and vibrational waves;  $Q_v (z,t)=\sqrt \rho x(z,t)$; $x$ is displacement of the vibrating particles,  $\rho $ is the medium density. With account for the energy and momentum conservation,
\begin{equation*} \label{eq2}
 \omega _l =\omega _s +\omega _v \left( {k_v } \right) ,\quad
 \vec {k}_l =\vec {k}_s \left( {\omega _s } \right)+\vec {k}_v,
\end{equation*}
one obtains the following equations for the slowly varying amplitudes in the approximation of the of first order of $Q$ in the polarization expansion:
\begin{eqnarray} \label{eq3}
&& \frac{\partial \varepsilon _l }{\partial z}+\frac{1}{v_l^{gr}
}\frac{\partial \varepsilon _l }{\partial t}=i\frac{\pi \omega _l^2 }{k_l
c^2}N\frac{\partial \alpha }{\partial Q}\varepsilon _s Q, \label{eq31}\nonumber \\
&& \frac{\partial \varepsilon _s }{\partial z}+\frac{1}{v_s^{gr}
}\frac{\partial \varepsilon _s }{\partial t}=i\frac{\pi \omega _s^2 }{k_s c^2}N\frac{\partial \alpha }{\partial Q}\varepsilon _l Q^*,\label{eq32}
\\
&& \frac{\partial Q}{\partial z}+\frac{1}{v_v^{gr}}\frac{\partial Q}{\partial
t}+\frac{Q}{\tau v_v^{gr}}=i\frac{1}{4\omega _v v_v^{gr}}N\frac{\partial \alpha }{\partial
Q}\varepsilon _l \varepsilon _s^* .\label{eq33}
\end{eqnarray}
Here, $v_{l,s,v}^{gr} $ are the projections of the group velocities of the fundamental, Stokes and vibration waves on the z-axis,  $N$ is the number density of the vibrating molecules, $\alpha $ is the molecule polarizability, $\tau $ is phonon lifetime, $\omega _0 $ is phonon frequency for $k_v =0$.
The dispersion $\omega _v(k_v)$ can be approximated as  \cite{ShB}
\begin{equation*} \label{eq4}
\omega _v =\sqrt {\omega _0^2 -\beta k_v^2 }.
\end{equation*}
Then,  in the vicinity of  $k_v=0 $, velocity $v_v^{gr}$  is given by:
\begin{equation*}
\label{eq5}
v_v^{gr} =-\beta {k_v }/{\omega _v }=-{\beta }/{v_v^{ph}},
\end{equation*}
where $v_v^{ph} $ is the projection of the phase velocity  of the vibrational wave on the z-axis and $\beta $ is the  dispersion parameter for the given crystal.

For the sake of clarity,  the continuous wave case and the approximation of the constant  field $E_l $ is considered. The latter is appropriate for the relatively week Stokes and vibrational waves. Then Eqs.~(\ref{eq32})-(\ref{eq33}) take the form:
\begin{equation*} \label{eq6}
{dQ}/{dz}=-ig_1\varepsilon _s^* -Q/({\tau v_v^{gr}}), \/
 {d\varepsilon _s }/{dz}=ig_2 Q^*,
\end{equation*}
where  $g_1=-N(\partial\/\alpha /\partial Q) \varepsilon _l/(4\omega _v v_v^{gr})$ and $g_2 =(\pi\omega _s^2/k_s c^2)
N (\partial \alpha /\partial Q) \varepsilon _l $.
In the case of Fig.~\ref{f1}(a), Eqs.~(\ref{eq6}) exhibit \emph{three fundamental differences} as compared with TWM of co-propagating waves in ordinary materials: an opposite sign with $g_1$ which stems from  $v_v^{gr}<0$, an opposite sign with $Q/({\tau v_v^{gr}})$ because  the phonon flow is against the $z$-axis, and the boundary conditions for $Q$ to be defined at $z=L$, i.e. at the opposite edge of the slab as compared to that for $\varepsilon _s$. This leads to \emph{fundamental changes} in their solutions and, consequently, in  the spatial and output behavior of the Stokes signal. Alternatively, in the given constant $\varepsilon _l$ approximation, the equations become identical and the behavior standard for the case of Fig.~\ref{f1}(b).
\section{Extraordinary properties of coherent energy conversion from fundamental to Stokes electromagnetic waves}
The solution to Eqs.~(\ref{eq6}) is found in the form:
\begin{equation}
\label{eq7}
 Q^*=A_1 e^{\beta _1 z'}+A_2 e^{\beta _2 z'  },
 \varepsilon _s=A_3 e^{\beta _1 z'}+A_4 e^{\beta _2 z'  },
\end{equation}
where $\beta _{1,2} =1 \mp iR$, $R=\sqrt{g_1^* g_2 l_p^2 -1} $,
$z'=z/l_{p}$,  $l_{p} =-2v_v^{gr} \tau $. The amplitudes $A_{1-4} $ and their relationships are determined by the boundary conditions. Transmission factors for co-propagating, $T_s^{\upuparrows}(z)$, and counter-propagating ($g_2<0$), $T_s^{\uparrow\downarrow}(z)$,  fundamental and Stokes waves are found as
\begin{eqnarray}
T_s^{\upuparrows}=\left|\frac{e^{z'}\left\{R\cos \left[ R\left(L'-z'\right)\right]+\sin \left[ R\left(L'-z'\right)\right]\right\}}{R\cos \left(RL'\right)+\sin \left(RL'\right)} \right|^2, \label{eq12}\\
T_s^{\uparrow\downarrow}=\left| \left\{\beta_1 e^{[\beta_2 (L'-z')]}-
 \beta_2 e^{[\beta_1 (L'-z')]}\right\}/{2R} \right|^2, \label{eq14}
\end{eqnarray}
where  $L'=L/l_{p}$, $T_s^{\upuparrows}=\left|\varepsilon_s(z)/\varepsilon _s(z=0)\right|^2$ and  $T_s^{\uparrow\downarrow}=\left|\varepsilon _s(z)/\varepsilon _s(z=L)\right|^2$.
Equations (\ref{eq12}) and (\ref{eq14}) display spatial distributions which are \emph{controlled} by the field $\varepsilon _l$ and are in a strict contrast to each other. It is explicitly seen for the ultimate  loss-free case ($l_p\rightarrow\infty$). Then
\begin{eqnarray}
 T_s^{\upuparrows}(z=L)&\rightarrow& 1/\cos^{2}(g L), \label{c}\\
T_s^{\uparrow\downarrow}(z=0)&\rightarrow& [\exp(2|g|L)]/4,\label{e}
\end{eqnarray}
where $g=\sqrt{g_1^* g_2}$. Equation~(\ref{c}) depicts a series of sharp giant  resonance enhancements of the output signal for  $g\rightarrow (2j+1)\pi/2L, (j= 0, 1, 2...)$. On the contrary, the coupling scheme of Fig.~\ref{f1}(b) is equivalent to scattering  on acoustic phonons and on optical phonons with positive group velocity. Correspondingly, Eq.~(\ref{e}) displays typical exponential growth with no resonances with respect to intensity of the fundamental control field. In general case, the denominator in Eq.~(\ref{eq12}) can be turned to zero if  $g^2 l_p^2>1$. The threshold value of intensity of the control field is
\begin{equation}
\label{eq11}
I_{\min } =\left({cn_s\lambda_{s0}\omega_v}/{8\pi^3l_p\tau }\right)\left| {N\partial \alpha /\partial Q}  \right|^{-2},
\end{equation}
where $\lambda_{s0}$ is Stokes wavelength in the vacuum.
\begin{figure}[htbp]
\begin{center}
\includegraphics[width=.98\columnwidth]{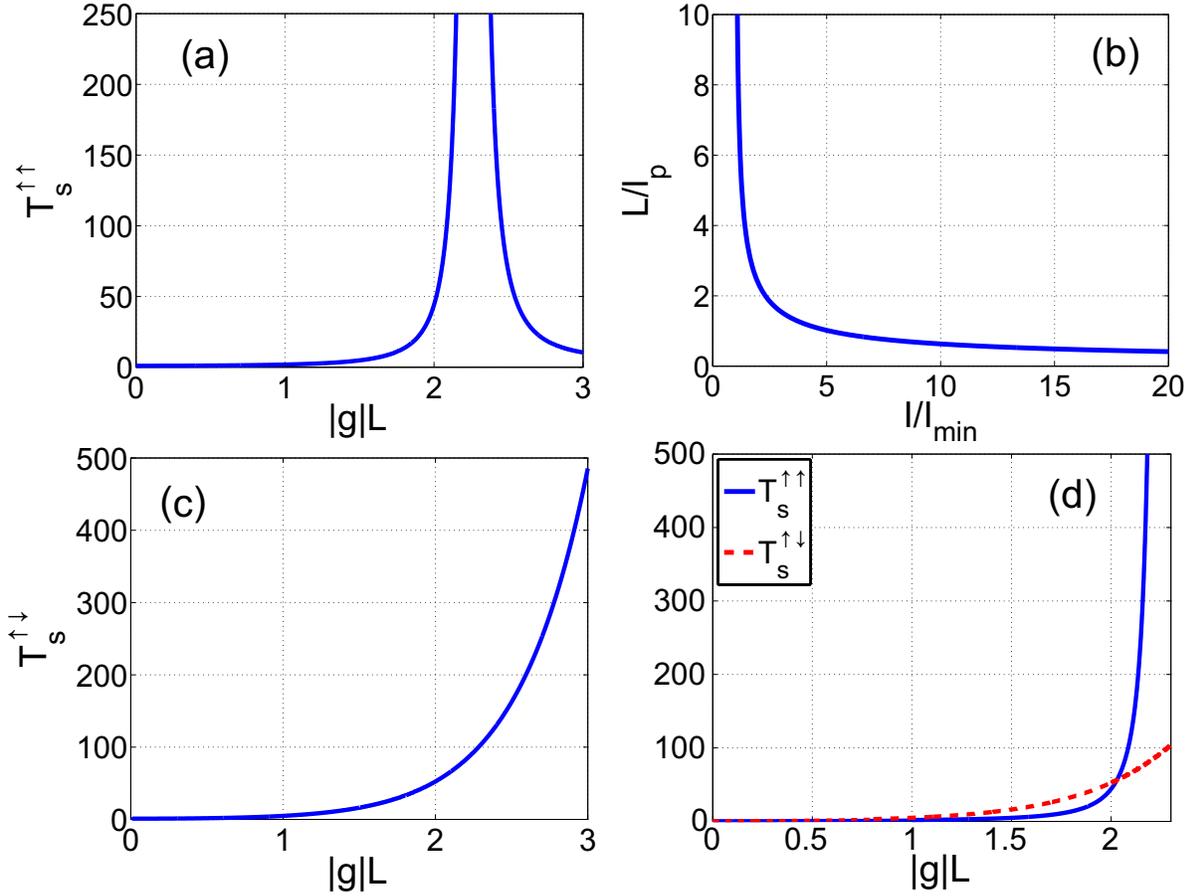}
\caption{\label{f2} (a) Transmission of the Stokes wave $T_s^{\upuparrows}(z=L)$ vs intensity of the fundamental control field in the vicinity of the first resonance (co-propagating geometry). (b) The crystal thickness corresponding to the first output resonance for the Stokes wave vs intensity of the control field for the co-propagating geometry; $I_{\min } $ is the threshold intensity. (c) Transmission $T_s^{\uparrow\downarrow}(z=0)$ vs intensity of the fundamental beam (contra-propagating geometry). (d) Comparison of the output intensities of the Stokes wave vs intensity of the control field for co- and contra-propagating coupling geometries.}
\end{center}
\end{figure}
For a given intensity of the control field $I_l>I_{min}$, the crystal thickness corresponding to the first resonance is
\begin{equation*}
\label{eq13}
L'=\left[{\pi -\tan^{-1} \left( R \right)}\right]/{R}.
\end{equation*}
Figure \ref{f2}(a) depicts transmission in the vicinity of the first resonance and Fig.~\ref{f2}(b) -- its position. In the resonance, $T_s^{\upuparrows}\rightarrow\infty$, which is due to the approximation of constant control field. Conversion of the control field to the Stokes one and excited molecule vibrations would lead to saturation of the control field and thus limits the maximum achievable amplification. Strong amplification in the maximums indicates the possibility of self-oscillations and thus creation of \emph{mirrorless} optical parametrical oscillator with unparalleled properties.

In the case of Fig.~\ref{f1}(b),  $k_s<0$ and the denominator in Eq.~(\ref{eq14}) cannot equal to zero. As outlined, this results in exponential spatial dependence with no resonances depicted in Fig.~\ref{f2}(c). Figure~\ref{f2}(d) shows that, in the vicinity of the resonance, three-wave coupling of waves with co-directed wave vectors and contra-directed energy flows of vibrational and Stokes waves  provides for essentially higher efficiency of coherent energy conversion  than in the standard  schemes. Figures~\ref{f2}(b, d) indicate the possibility to \emph{fit in} the effective conversion length within the crystal of a given thickness and to significantly \emph{concentrate} the generated Stokes field nearby its output facet.

Estimations made for the model, which is characteristic for the diamond crystal, $\omega_v=1332$ cm$^{-1}$ and vibrational transition width $(c\tau)^{-1}=1.56$ cm$^{-1}$ \cite{G,Ch,An}, show that the required excitation intensities are above the typical crystal breakdown threshold in the continuous wave regime. However, the cross-section of the Raman scattering is inverse proportional to the squared frequency offset from the intermediate single-photon resonance. Therefore, the threshold intensity $I_{min}$ given by  Eq.~(\ref{eq11}) can be reduced by seven to eight orders compared to its off-resonance values by approaching  such a resonance. The required intensity is also achievable with the commercially available pulsed lasers, especially  for the pulse length shorter than the phonon relaxation rate.
\section{Conclusions}
To conclude, we propose extraordinary coherent energy conversion process attributed to  three-wave mixing of ordinary and backward waves. It rests on the excitation of optical phonons with negative dispersion and can be implemented in the routinely fabricated  crystals extensively studied by the means of Raman spectroscopy. The investigated features can be employed for creation of a family of the unique nonlinear-optical photonic devices, such as frequency-tunable switches, filters, amplifiers and miniature mirrorless optical parametric oscillators. Earlier, such a possibility was proposed based on negative index metamaterials which are much more difficult to fabricate and are inherently strongly lossy. The revealed features stem fundamentally from the opposite directions of the phase velocity and the energy flow, which is the characteristic property of optical phonons having negative dispersion, $\partial \omega(k)/\partial k < 0$.

\section*{Acknowledgments}
This work was supported by the Russian Federal Program on Science, Education and Innovation under Grant No 2010-121-102-018,  by the Presidium of the Russian Academy of Sciences under Grant No 27.1, by the Siberian Division of the Russian Academy of Sciences under Integration Project No~5 and by the US National Science Foundation under Grant  ECCS-1028353.


\begin{thebibliography}{99}
\bibitem{Sh} V. M. Shalaev, ``Optical negative-index metamaterials,''
Nat. Photonics \textbf{1}, 41--48 (2007).

\bibitem{SM} C. M. Soukoulis and M. Wegener,
Optical metamaterials — more bulky and less lossy,
Science \textbf{330}, 1633--1634 (2011).

\bibitem{Zh} N. I. Zheludev,
A roadmap for metamaterials,
OPN \textbf{22}, 30--35 (20011).

\bibitem{OL} A. K. Popov and V. M. Shalaev,
``Compensating losses in negative-index metamaterials by optical parametric amplification,''
Opt. Lett. \textbf{31}, 2169--2171 (2006).

\bibitem{EPJD}	A. K. Popov, ``Nonlinear optics of backward waves and extraordinary features of plasmonic nonlinear-optical microdevices,''  {Eur. Phys. J. D } \textbf{58}, 263--274  (2010) (topical issue on {Laser Dynamics and Nonlinear Photonics}).

\bibitem{LPL} A. K.	Popov, V. V. Slabko and V. M. Shalaev,
Second harmonic generation in left-handed metamaterials,
Laser Phys. Lett \textbf{3}, 293--297(2006).
\bibitem{APB1} A. K. Popov and V. M. Shalaev,
``Negative-index metamaterials:
second-harmonic generation, Manley-Rowe relations and parametric
amplifications,''
Appl. Phys. B \textbf{84}, 131--137 (2006).

\bibitem{Ma} L. I. Mandelstam,
Group velocity in a crystall lattice,
 ZhETF \textbf{15}, 475--478 (1945).

\bibitem{ShB} 	Y. R. Shen  and N. Bloembergen,
Theory of stimulated brillouin and raman scattering,
Phys. Rev. \textbf{137}, A1787--A1805 (1965).

\bibitem{Boy} R. W. Boyd, Nonlinear Optics, Third Edition (Amsterdam: Academic Press, 2008.


\bibitem{Bob} D. L. Bobroff,
Coupled-modes analysis of the phonon-photon parametric backward-wave oscillator,
J. Appl. Phys. \textbf{36}, 1760--1769 (1965).

\bibitem{Kh} J. B. Khurgin,
``Mirrorless magic,"
Nat. Photonics \textbf{1}, 446--448, (2007).

\bibitem{G} V. S. Gorelik,	Contemporary problems of Raman spectroscopy, Moscow, Nauka Publishing Co., 1978 (in Russian), pp. 28-47.
\bibitem{An}E. Anastassakis,  S. Iwasa and E. Burstein,    Electric-Field-Induced Infrared Absorption In Diamond,
    Phys. Rev. Lett. \textbf{17}, 1051--1054 (1966).

\bibitem{Ch} Y.	Chen and J. D. Lee,
Determining material constants in micromorphic theory through phonon dispersion relations,
International J. of Engineering Science \textbf{41}, 871--886 (2003).

\end{thebibliography}
\end{document}